\documentclass[twocolumn,10pt]{asme2ej}
\usepackage{flushend}
\usepackage{graphicx} 
\usepackage{xcolor} 
\usepackage{MazzoleniNotation} 
\usepackage{hyperref} 
\usepackage{subfig}    
\usepackage{makecell}
\usepackage{float} 
\usepackage{color,soul} 
\graphicspath{{./figure/}}
\usepackage{array}
\newcolumntype{P}[1]{>{\centering\arraybackslash}p{#1}}

\title{Combined Plant and Controller Optimization of an Underwater Energy Harvesting Kite System}

\author{Kartik Naik
    \affiliation{
	PhD candidate\\
	Department of Mechanical and \\
	Aerospace Engineering\\
	North Carolina State University\\
	Raleigh, NC 27606\\
    Email: knaik@ncsu.edu
    }
}
\author{Sumedh Beknalkar
    \affiliation{
	PhD \textcolor{black}{student}\\
	Department of Mechanical and \\
	Aerospace Engineering\\
	North Carolina State University\\
	Raleigh, NC 27606\\
    Email: sbeknal@ncsu.edu
    }
}

\author{\textcolor{black}{James Reed}
    \affiliation{ \textcolor{black}{
	PhD student}\\
	\textcolor{black}{Department of Mechanical and} \\
	\textcolor{black}{Aerospace Engineering}\\
	\textcolor{black}{North Carolina State University}\\
	\textcolor{black}{Raleigh, NC 27606}\\
    \textcolor{black}{Email: jcreed2@ncsu.edu}
    }
}
\author{Andre Mazzoleni
    \affiliation{
	Professor\\
	Department of Mechanical and \\
	Aerospace Engineering\\
	North Carolina State University\\
	Raleigh, NC 27606\\
    Email: apmazzol@ncsu.edu
    }
}
\author{Hosam Fathy
    \affiliation{
	Professor\\
	Department of Mechanical \\
	Engineering\\
	University of Maryland\\
	Raleigh, NC 27606\\
    Email: hfathy@umd.edu
    }
}
\author{Chris Vermillion\thanks{Address all correspondence for other issues to this author.}
    \affiliation{Associate Professor\\ 
    Department of Mechanical and \\
	Aerospace Engineering\\
	North Carolina State University\\
	Raleigh, NC 27606\\
    Email: cvermil@ncsu.edu
    }
}

\begin{document}

\maketitle
%
\begin{abstract}
{\it \textbf{Abstract} - This paper presents the formulation and results for a control-aware optimization of the combined geometric and structural design of an energy-harvesting underwater kite. Because kite-based energy-harvesting systems, both airborne and underwater, possess strong coupling between closed-loop flight control, geometric design, and structural design, consideration of all three facets of the design within a single co-design framework is highly desirable. However, while prior literature has addressed one or two attributes of the design at a time, the present work constitutes the first comprehensive effort aimed at addressing all three. In particular, focusing on the goals of power maximization and mass minimization, we present a co-design formulation that fuses a geometric optimization tool, structural optimization tool, and closed-loop flight efficiency map. The resulting integrated co-design tool is used to address two mathematical optimization formulations that exhibit subtle differences: a Pareto optimal formulation and a dual-objective formulation that focuses on a weighted power-to-mass ratio as the techno-economic metric of merit. Based on the resulting geometric and structural designs, using a medium-fidelity closed-loop simulation tool, the proposed formulation is shown to achieve more than three times the power-to-mass ratio of a previously published, un-optimized benchmark design.}
\end{abstract}
%
\section{Introduction}

Marine hydrokinetic (MHK) energy is a tremendous clean renewable energy source, with an estimated 163 TWh/year of usable ocean current  energy, and  334 TWh/year of usable tidal energy in the United States alone \cite{haas2013assessment}. While such a resource can power millions of homes or aid in powering so-called ``Blue Economy" devices (e.g., offshore research platforms, autonomous underwater vehicles), harvesting it can be challenging. 
Fixed ocean turbines are accompanied with challenges of size of system (for example, the required size for a fixed ocean turbine operating in a 1 m/s flow ocean environment is the same as the size of a towered wind turbine operating in a 10 m/s ground environment, making the required size per unit power similar in each case, considering densities) and limited viable locations for such extraction of such resources. \\
\indent One class of system that is designed specifically to overcome the aforementioned challenge of size and also enable economic energy extraction in lower-flow sites is a MHK kite. An MHK kite is able to harvest ocean current energy by flying in a \emph{cross-current} figure-8 or elliptical path in the flow environment. The power generation takes place either with an on-board rotor (fly-gen), or through generator at a ground station (ground-gen) where cyclic spooling (high-tension spool-out motions and low-tension spool-in motions) to generate power \textcolor{black}{\cite{ghasemi2018computational,olinger2015hydrokinetic}}. In this work, the ground-gen system is considered, where cross-current flight is accomplished through figure-8 motions. \\
\indent The bulk of the literature on MHK kites (including but not limited to \cite{reed2020optimal,ghasemi2015computational,li2015control,siddiqui2020lab}) and their sister technology, airborne wind energy (AWE) systems (including but not limited to \textcolor{black}{\cite{Vermillion_survey,williams2008optimal,Fagiano_ACC_AWE,rapp2019modular,fechner2013model,ranneberg2018fast})}, has concentrated on dynamic modelling and controller design. While this focus is warranted given the control challenges associated with cross-current flight of such a complex dynamic system, there are a host of plant design intricacies that need to be considered in order to optimally harvest the MHK resource. These are associated with (i) geometric kite design for hydrodynamically efficient cross-current motion and (ii) structural design to support the hydrodynamic loads incurred during flight. While some of the AWE system literature, including \cite{aull2020design,candade2020aero,candade2020structural,fasel2017aerostructural}, address the plant optimization with structural consideration, they do not consider the closed-loop flight dynamics of cross-current flight in their optimization. 

Our preliminary work in \cite{naikfused} (which the present journal paper serves as a significant extension of) shows, in fact, that the achievable closed-loop flight efficiency of a kite-based system, measured relative to the quasi-static predictions of Loyd in \cite{loyd1980crosswind}, varies considerably with geometric design parameters, which are in turn coupled with structural parameters. In light of these facts, our initial work in \cite{naikfused} presented a fused geometric, structural and control co-design framework that maximized power-to-mass ratio of the kite subject to performance, geometric and structural constraints. The closed-loop flight performance (detailed in Section 4) was mapped to flight performance decision variables to act as a \emph{control proxy function}. In the formulation, the overall optimization was divided into three modules: 
\begin{enumerate}
    \item A steady flight optimization tool (SFOT), which selects wing and stabilizer properties to minimize displaced volume (surrogate for mass), subject to geometric and performance constraints; 
    \item A structural wing design tool (SWDT), which selects spar and skin properties to minimize wing mass, subject to wing tip deflection and buoyancy considerations; 
    \item A structural fuselage design tool (SFDT), which selects the fuselage thickness to minimize fuselage mass, subject  to  hoop  stress,  sheer  stress,  and  buckling  constraints. 
\end{enumerate}

The generated power in the aforementioned SFOT was estimated using a control proxy function (CPF), which mapped flight efficiencies to performance decision variables. This efficiency characterizes the difference between actual power generated in simulation and the theoretical optimum steady cross-current flight power, based on quasi-static assumptions made in \cite{loyd1980crosswind}. The CPF was used as a surrogate for full dynamic simulations within the optimization framework of \cite{naikfused}, which drastically reduced time for carrying out the optimization, while still achieving a 22$\%$ improvement in the kite's performance relative to an assumption of plant-independent flight efficiency. 


While the framework in \cite{naikfused} simultaneously considers the kite's geometry, structure, and closed-loop flight efficiency, it does so in a specialized manner consisting of a ``nested sequential" formulation (where a sequential wing geometric and structural optimization process is nested within a loop that iterates on geometric fuselage parameters), focusing specifically on the minimization of mass subject to a fixed power requirement. As we will illustrate in the present work, while the nested sequential approach of \cite{naikfused} efficiently converges upon an \emph{improved} kite design, it can miss the true global optimum due to the sequential nature of the wing geometry and structural optimization. \textcolor{black}{Furthermore, while the CPF used in \cite{naikfused} was based on judicious tuning of control parameters for each combination of plant decision variables, the control parameters were not in fact optimized.} 

\textcolor{black}{To remedy the limitations of the nested sequential approach of \cite{naikfused}, the present paper details a fully nested formulation that is nevertheless computationally efficient. To remedy the non-optimality of the CPF from \cite{naikfused}, the present paper leverages an economic iterative learning control (ILC) strategy first presented in \cite{cobb2022flexible}, which iteratively updates the control parameters that define the kite's flight path, in pursuit of maximizing the lap-averaged power.} Furthermore, using the fully nested formulation, the present paper examines the Pareto front traced through the variation of the fixed power requirement in a single-objective mass minimization. This investigation includes the presentation of the Pareto front itself, along with the selected physical designs (in terms of geometry and wing structure) that correspond to select points along the Pareto front. Finally, the paper presents the results of a dual-objective formulation where the performance metric of interest is a weighted power-to-mass ratio, again utilizing the fully nested framework. The dual-objective formulation provides an intuitive mechanism of directly introducing techno-economic measures of interest into the optimization problem, while producing a result that traces the convex portions of the aforementioned Pareto front. 

The remainder of the paper is organized as follows: Section 2 summarizes the overall physical system. Section 3 reviews the Pareto optimal and dual-objective problems to be addressed. Section 4 reviews the constituent tools developed to address each component of the optimization problem (the geometry, structure, and closed-loop flight efficiency). Section 5 describes the nested-sequential, fully nested, and simultaneous topologies for fusing the aforementioned tools into an overall optimization framework. Finally, in Section 6, simulation results are presented, along with a Pareto front and corresponding designs along that Pareto front that illustrate the trade space between power maximization and mass minimization.

\section{System Formulation}
\begin{figure*}[thpb] 
\begin{center}
  	\includegraphics[width = 0.85\linewidth]{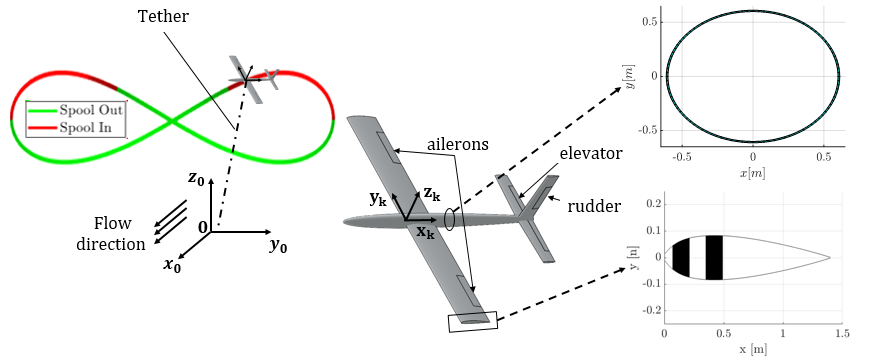}
	\caption{\textcolor{black}{(Left) The ocean kite flying in a figure-8 path, with the inertial frame of reference at the winch shown as $0$-frame and the flow of the current being along the $\mathbf{{x_0}}$ axis. (Right) The outer mold line of the kite, along with the control surfaces used to control the orientation of the kite, and the body-fixed frame of reference $\mathbf{k}$ are shown. The plots adjacent to it show a sample output of the structural tools used to design the fuselage shell (top) and the wing structure (bottom) respectively.}
	\label{fig: system}}  
\end{center}
\end{figure*}

Several embodiments of the physical kite system considered in this work, along with the coordinate system used in the dynamic model, are shown in Fig.\ref{fig: system}. The kite itself is a rigid glider with four hydrodynamic surfaces -- two ailerons, a rudder, and an elevator -- which generates net positive energy by spooling out under high tension and spooling in under low tension. Tension is modulated by varying the kite's angle of attack over a figure-8 flight path, utilizing the edges of the flight path for spool-in operation and the center of the path for spool-out operation. The structural wing elements of the kite include multiple spars as well as the wing skin, whereas the fuselage acts as a pressure vessel and beam, whose rigidity is provided through its wall thickness.

\begin{figure*}[thpb]
\begin{center}
  	\includegraphics[width = \linewidth]{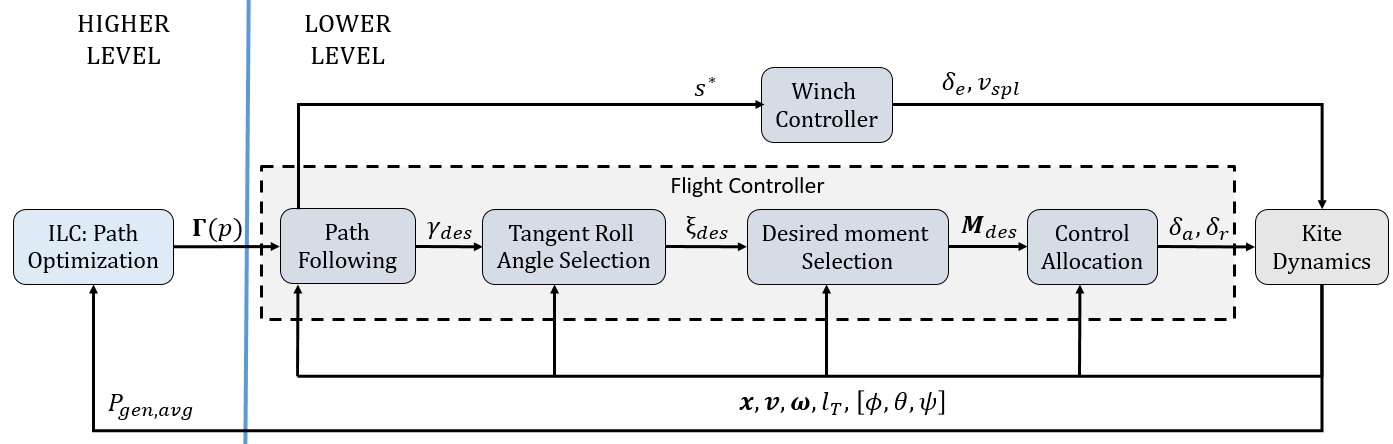}
	\caption{\textcolor{black}{Block diagram of the control system: The higher level iterative learning controller optimizes the path vector ($\Gamma(p)$) to maximize power generation. The lower level controller is responsible for path following (flight controller) and spooling operation (winch controller), where \textbf{\emph{x}} is the position vector, \textbf{\emph{v}} is the velocity vector, $\mathbf{\omega}$ is the kite angular velocity vector, $l_T$ is the un-spooled tether length, and $\left [ \phi, \theta, \psi  \right ]$ is the vector of Euler angles (roll, pitch and yaw respectively).}}
	\label{fig: control}  
\end{center}
\end{figure*}

\subsection{Plant Geometry}
As illustrated in Fig.\ref{fig: system}, the design of the kite considered in this work is similar to that of a glider -- it contains a starboard and port wing, a fuselage, and a horizontal and vertical stabilizer at the aft section of the kite. The kite's attitude is controlled through the actuation of three sets of control surfaces: ailerons, elevators and a rudder. The kite is connected to the ground station with a single tether that is used for data and power transmission, and is designed to handle the tensile loads generated during cross-current flight. \\
\indent To provide input data for the geometric design tool and kite dynamic model, each of the kite's hydrodynamic surfaces was characterized using XFLR \cite{drela2012xflr}. A parametric fit was obtained after running a batch analysis for wings with aspect ratios varying from 4 to 25. The resulting hydrodynamic coefficients were fit to the following standard lift and drag equations \cite{torenbeek2013synthesis}: 
\begin{flalign}
C_L& = \frac{2\pi\gamma}{1+\frac{2 \gamma}{e_L AR}}\cdot \alpha + C_{L,0}, \label{eqn:lift}
\end{flalign}
\begin{flalign}
C_D& = (\frac{1}{\pi e_D AR} + K_{\mathrm{visc}})\cdot(C_L-C_{L,x})^2 + C_{D,0}.
\label{eqn:drag}
\end{flalign}
\noindent where $\alpha$ is the angle of attack, $C_{L,0}$ is the lift at $\alpha = 0$, $\gamma$ is the foil lift curve multiplier, $e_L$ is the Oswald lift efficiency, $e_D$ is the Oswald drag efficiency, $K_{\mathrm{visc}}$ is the viscous drag coefficient factor, $C_{L,x}$ is the lift at minimum drag (non-zero for cambered airfoils), and $C_{D,0}$ is the drag at zero lift. 

\subsection{Plant Structural Design and Characterization}
The structural design tools, SWDT and SFDT, are used to design support structures for the wing and fuselage respectively. Both tools minimize structural mass (of the wing and fuselage, respectively) subject to constraints based on the loading and material properties of the respective structural elements. Al 6061 was selected for the structural elements due to its high specific modulus \cite{wenzelburger2010manufacturing} (ratio of the elastic modulus to material density), the relevant property to design a stiffness-driven structure like a wing. 
\subsubsection{Wing structure}
The wing is assumed to be rectangular in planform, thus having a uniform cross-section. The section consists of spars (ranging from 1 to 3 in number), and a shell. The structural decision variables are the number of spars $N_{sp}$, the thickness of each spar $t_{sp}$ and the thickness of the wing shell $t_{s,w}$. \textcolor{black}{The chord-wise location, of each spar is a user-defined constant and depends on the number of spars present, as noted in Table \ref{tab: spar loc}.}  The shell and the outer edges of the spar take on the shape of a user-defined airfoil (\textcolor{black}{NACA 2412} \cite{tools2015naca} is considered for this work) as shown in Fig.\ref{fig: system}. The structural mass is computed as:  
\begin{flalign}
m_\mathrm{{wing}}(\mathrm{c},\mathbf{u_{WD}}) = \rho_m s A_\mathrm{{wing}}(\mathrm{c},\mathbf{u_{WD}})
\end{flalign}

where $c$ is the wing chord, $s$ is the wing span, $\rho_m$ is the material density and $A_{wing}$ is computed by integrating the curves shown on the bottom right corner of Fig. \ref{fig: system}.

\begin{table}[h]
\centering
\caption{\textcolor{black}{Chord-wise spar location (distance from the wing's leading edge)}}
\begin{tabular}{>{\color{black}}c| >{\color{black}}c >{\color{black}}c >{\color{black}}c}
         & \multicolumn{3}{>{\color{black}}c}{spar location [$\%$ of $c$]}                         \\ \cline{2-4} 
$N_{sp}$ & \multicolumn{1}{>{\color{black}}c|}{spar 1} & \multicolumn{1}{>{\color{black}}c|}{spar 2} & spar 3 \\ \hline \hline
1        & \multicolumn{1}{>{\color{black}}c|}{-}   & \multicolumn{1}{>{\color{black}}c|}{25}  & -   \\
2        & \multicolumn{1}{>{\color{black}}c|}{10}  & \multicolumn{1}{>{\color{black}}c|}{40}  & -   \\
3        & \multicolumn{1}{>{\color{black}}c|}{15}  & \multicolumn{1}{>{\color{black}}c|}{30}  & 60  \\ \hline
\end{tabular}
\label{tab: spar loc}
\end{table}

The physicality of the system is modeled using an inequality constraint on the maximum allowable wingtip deflection, $\delta_{max}$, as well as bounds on decision variables, which are modeled as a fraction of the wing chord and thickness to scale changes in wing chord during the optimization. The maximum wing tip deflection is calculated assuming a point load acts on the centroid of the cantilevered wing \cite{hearn1997mechanics}: 
\begin{flalign}
\delta_\mathrm{{max}} = \frac{F_\mathrm{{wing}}}{6 E I_\mathrm{{req}}}(1.5S - a) a^2
\end{flalign}
where $E$ is the Young's modulus, $s$ is the wing span, and $a$ is the span-wise location of the centroid of the wing. As the kite is designed to be neutrally buoyant, the gravity and buoyancy forces cancel out, and we assume that the net bending force, $F_{\mathrm{wing}}$, is the lift force on each wing (half the total lift force). The inequality constraint on wingtip deflection can be translated to an inequality constraint on the resulting moments of inertia (as formulated in \eqref{eqn:Ireq}), requiring that the moment of inertia of each wing , $I_{\mathrm{wing}}$, be greater than the calculated required moment of inertia, $I_{\mathrm{req}}$. 

\subsubsection{Fuselage structure}
The fuselage is modeled as a thin cylindrical pressure vessel. Following this, (i) hoop stress due to the net pressure difference, (ii) shear stress caused due to hydrodynamic forces of the lifting surfaces, and (iii) buckling due to the lift forces of the wing and the horizontal stabilizer, are modeled as inequality constraints with a safety factor (detailed in section \ref{sec: SFDT}). 


\textcolor{black}{\subsection{Dynamic Model}}
\textcolor{black}{
A dynamic model of the integrated kite and tether, detailed in \cite{reed2020optimal,abney2022autonomous}, was used to simulate the flight performance of the kite in this work. The instantaneous mechanical power generated by the kite is given by: 
\begin{flalign}
P_{\mathrm{gen}}(t) = \left \| \mathbf{F_{thr}}(t)) \right \|v_{\mathrm{spl}}(t)&
\end{flalign}
\noindent where $v_{\mathrm{spl}}$ is the spooling speed and $ \mathbf{F_{thr}}$ is the tether tension.} 

\textcolor{black}{The kite, modeled as a rigid lifting body, experiences external forces and moments due to gravity, buoyancy and hydrodynamics. The tether dynamics are captured using a lumped mass model, each link of which is characterized as a non-compressible spring-damper as detailed in \cite{vermillion2013model}. The kite dynamics are reformulated to account for added mass effects using the method described in \cite{fossen2011handbook}, based on strip theory. The following sections summarize the kite and tether models used.}

\textcolor{black}{\subsubsection{Kite model}}
\textcolor{black}{As illustrated in Fig. \ref{fig: system}, the kite's reference frame is defined by three body-fixed orthonormal unit vectors, $\mathbf{x_k}$, $\mathbf{y_k}$, and $\mathbf{z_k}$, whose origin is at the intersection of the kite's vertical plane of symmetry and the leading edge of the wing. The state variables of the 6-degree of freedom flight dynamic model are the kite's position, orientation and their respective rates of change. The kite's nonlinear equations of motion are formulated as:}
\textcolor{black}{\begin{equation}
    \dot{\mathbf{v}}_r = \mathrm{M_k}^{-1} \left(\mathbf{\tau}(\mathbf{v}_r)-\mathbf{C}\left(\mathbf{v}_r\right)\mathbf{v}_r\right),
\end{equation}
where 
\begin{equation}
    \mathbf{v}_r = \begin{bmatrix}
    u_\mathrm{kite} - u_f\\
    v_\mathrm{kite} - v_f\\
    w_\mathrm{kite} - w_f\\
    p\\
    q\\
    r\\
    \end{bmatrix}. \label{eq:dynVar}
\end{equation}}
\textcolor{black}{In \eqref{eq:dynVar}, $u_{\mathrm{kite}}$, $v_{\mathrm{kite}}$, $w_{\mathrm{kite}}$, $u_{f}$, $v_{f}$, and $w_{f}$ are, respectively the body frame velocity components of the kite and flow, where $u$, $v$, and $w$ correspond to the velocities along the kite $\mathbf{x_k}$, $\mathbf{y_k}$, and $\mathbf{z_k}$ axes, respectively; $p$, $q$, and $r$ are the angular rates about the kite's body axes in the kite body-fixed frame.} 
\par
\textcolor{black}{The vector $\mathbf{\tau}$ comprises of the net external forces and the moments acting on the kite: 
\begin{equation}
    \bold{\tau} = \begin{bmatrix}
    \mathbf{{F}_{Net,k}}\\
    \mathbf{{M}_{Net,k}}\\
    \end{bmatrix}
\end{equation}}

\noindent \textcolor{black}{where $\mathbf{{F}_{Net,k}}$,$\mathbf{{M}_{Net,k}} \in  \mathbb{R}^{3}$ are the vectors containing the external forces and moments acting on the kite respectively. The vector $\mathbf{\tau}$
includes gravitational, buoyancy, tether and hydrodynamic forces (dependent on $\mathbf{v}_r$) and associated moments. Additionally, $\mathbf{C}(\mathbf{v}_r) \in \mathbb{R}^{6 \times 6}$ accounts for the Coriolis terms arising from rigid body and added mass effects. Lastly, $\mathrm{M_k}\in \mathbb{R}^{6 \times 6}$ is the mass matrix of the kite, which comprises mass and inertia terms, coupling terms arising from the kite's center mass position being displaced from the body-fixed origin, and added mass terms.}

\textcolor{black}{The kite is subjected to forces and moments arising from five hydrodynamic surfaces (a port wing, starboard wing, horizontal stabilizer, vertical stabilizer and fuselage), buoyancy, gravity and the tether dynamics. The net force vector is computed as:}

\textcolor{black}{\begin{align}
\begin{split}\label{eqn:translationalForces}
    \mathbf{{F}_{Net,k}} =& \mathbf{{F}_{thr}} + \left( V_{kite}\rho_w - m_{kite} \right)g\mathbf{{z}_{o}}\\
    & + \frac{1}{2} \rho_w S_{ref} \sum_{i=1}^{5}\| \mathbf{{v}_{a_i}}\|^2 \left( C_{L,i}\mathbf{{u}_{L,i}} + C_{D,i}\mathbf{{u}_{D,i}}\right)
   \end{split}
\end{align}}
\noindent \textcolor{black}{where $\mathbf{{F}_{thr}}$ is the tether tension vector, $V_{kite}$ and $m_{kite}$ are the volume and mass of the kite, $\rho_w$ is the fluid density, $g$ is the gravitational acceleration, $S_{ref}$ is the kite's reference area (referenced to the wing planform area), $\mathbf{{v}_{a_i}}$ is the apparent velocity of each hydrodynamic surface where index $i$ references each of the five hydrodynamic surfaces, and $C_{L,i}$ and $C_{D,i}$ are the coefficients of lift and drag, respectively, for each hydrodynamic surface. The second and third term in equation \eqref{eqn:translationalForces} describe the net buoyant and hydrodynamic forces.} 

\textcolor{black}{The net moment vector is calculated as the sum of the cross products of the individual hydrodynamic forces and the associated moment arm, $\mathbf{r_{a_i}}$: 
 \begin{align}\label{eqn:netMoment}
     \mathbf{{M}_{Net,k}} \!=&\!   \mathbf{r_{a_i}} \times\! \frac{1}{2} \rho_w S_{ref}\!\! \sum_{i=1}^{5}\| \mathbf{{v}_{a_i}}\|^2 \left( C_{L,i}\mathbf{{u}_{L,i}} + C_{D,i}\mathbf{{u}_{D,i}}\right)\! \\ 
     \nonumber+ &\mathbf{{r}_{ta}} \times\!\mathbf{{F}_{thr}}  +\mathbf{{r}_{cb}} \times\!V_{kite} \rho_{w}\mathbf{{z}_{o}}-\mathbf{{r}_{cg}} \times\!m_{kite} g\mathbf{{z}_{o}}
 \end{align}}
\noindent \textcolor{black}{where $\mathbf{{r}_{cb}}$ is the vector from the center of buoyancy to the leading edge of the wing, $\mathbf{{r}_{ta}}$ is the vector from the tether attachment point to the leading edge of the wing, and $\mathbf{{r}_{cg}}$ is the vector from the center of gravity to the leading edge of the wing. In equations \eqref{eqn:translationalForces} and \eqref{eqn:netMoment}, $\mathbf{{u}_{L,i}}$ and $\mathbf{{u}_{D,i}}$ represent the unit vectors describing the direction of the lift and drag forces at the at the $i^{\text{th}}$ hydrodynamic center.} 

\textcolor{black}{\subsubsection{Tether model}}
\textcolor{black}{The tether is modeled as a chain of non-compressive springs (links), connected with point masses (nodes), subject to buoyancy, gravity and drag forces. The subscript $c$ is used to denote each lumped mass, where $c = 1,2,\hdots,N_c$, and $N_c$ is the number of tether nodes. The net force acting on each tether node is given by:}
\textcolor{black}{\begin{equation}
\label{eq:tet}
\begin{split}
  \mathbf{{F}_{thr,c}} =&\frac{1}{2}\big( \left( \rho_w -\rho_{thr} \right)\pi r_{thr}^2 l_{T,c} g \mathbf{{z}_{o}} +\mathbf{{F}_{Ten,c}}\\
    &+ \frac{1}{2}\rho_w \|\mathbf{{v}_{app,thr,c}}\|^2 A_{p,thr,c}C_{D,thr,c}\frac{\mathbf{{v}_{app,thr,c}}}{\|\mathbf{{v}_{app,thr,c}\|}}\big)\\
\end{split}
\end{equation}}
\textcolor{black}{Here, \noindent $\rho_{thr}$ is the density of the tether, $r_{thr}$ is the radius of the tether, $l_{T,c}$ is the un-spooled tether length, $A_{p,thr,c}$ is the area of the tether projected in the direction of the apparent flow, $C_{D,thr,c}$ is the tether drag coefficient, $\mathbf{{v}_{app,thr,c}}$ is the apparent flow speed at the center of the tether link, and $\mathbf{{F}_{Ten,c}}$ is the nonlinear spring-damper force, which is equal to zero if $\|\mathbf{r_c}\| < l_{T,c}$. If $\|\mathbf{r_c}\| \geq l_{T,c}$, $\mathbf{{F}_{Ten,c}}$ is computed as:} 
\textcolor{black}{\begin{equation}
\begin{split}
\label{eq:tet1}
    \mathbf{{F}_{Ten,c}} = 
    &\frac{1}{2}\bigg(-E_{y,c}\frac{\pi r_{thr}^2}{l_{T,c}} \left( \|{\bold{r_{c}}}\|-l_{T,c}\right)\\&\quad-2\zeta_c \sqrt{E_{y,c}\frac{\pi r_{thr}^2}{l_{T,c}} m_c}\frac{d}{dt}\|{\bold{r_{c}}}\| \bigg)\frac{{\bold{r_{c}}}}{\|{\bold{r_{c}}}\|}
\end{split}
\end{equation}}
\textcolor{black}{\noindent where $E_{y,c}$ is the Young's modulus, $\zeta_{c}$ is the non-dimensional damping ratio, $m_{c}$ is the damping mass, and $\mathbf{r_{c}}$ is the vector from the origin to each lumped mass.}

\textcolor{black}{\subsection{Higher-level Iterative Learning Controller (ILC)}}

\textcolor{black}{The ILC algorithm used in this work is responsible for optimizing the path geometry for each kite design to generate a flight efficiency map. This algorithm is described in \cite{cobb2022flexible}, and a summary is included here for self-containment. The goal of this ILC algorithm is to maximize an economic objective, which for the kite system is the lap-averaged power production, by altering a set of parameters $\mathbf{b}$ that define the path's shape. To accomplish this, the algorithm performs two steps in sequence. They are: 
\begin{enumerate}
    \item a meta-model identification step, where the performance metric is estimated as a function of control parameters that describe the path shape followed by the kite using a recursive least squares fit;
    \item an flight parameter update, where the performance from previous figure-eight cycles is used to update the set of control parameters that define the kite's path using a gradient-based formulation with an added perturbation for persistent excitation. 
\end{enumerate}}

\begin{figure}[h] 
\begin{center}
  	\includegraphics[width = \columnwidth]{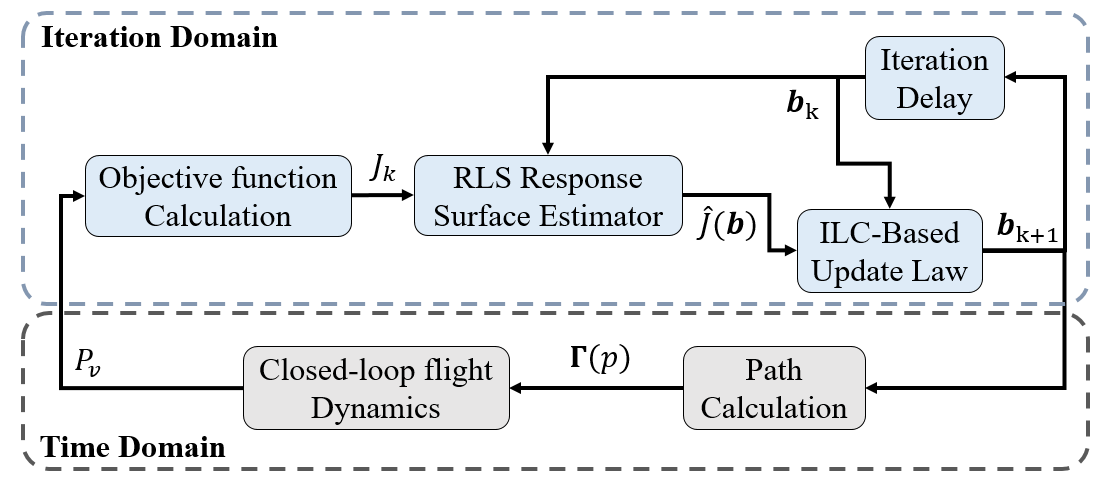}
	\caption{\textcolor{black}{Block diagram depicting the ILC-based flight path optimization. The variable $P_v$ in the figure is used to denote plant variables.} \label{fig: ilcBD}}  
\end{center}
\end{figure}
\color{black}
A block diagram of this process can be seen in Fig.\ref{fig: ilcBD}. The objective function being maximized is calculated as
\begin{equation}\label{eqn:PerformanceIndex}
J_j = \frac{1}{t_{e,j}-t_{s,j}}\int_{t_{s,j}}^{t_{e,j}}[P_{gen}(\tau) - k_w d(\tau)]d\tau,
\end{equation}
\noindent where $t_{s,j}$ and $t_{e,j}$ are the start and end times of the $k$-th iteration, where an iteration is defined as one figure-eight lap, $k_w$ is a scalar weight, and $d(t)$ is the penalty that describes our secondary design objectives.  In this work, we use $d(t)=\gamma_c(t)$ where $\gamma_c(t)$ is the interior angle to the path from the kite. This is a measure of how far from the path the kite is at each instant.

The basis parameters ($\mathbf{b}$) used in this work define the figure-8 path, which is based on the lemniscate of Booth as shown in \cite{rapp2019modular}. In spherical coordinates, the path is then described at any radius by the equations for azimuth and elevation angles, $\Phi_\Gamma$ and $\Theta_\Gamma$,
\begin{align}\label{eqn:LemniscateOfBooth}
    \Phi_\Gamma(p,\mathbf{b}) &= \frac{(\frac{b_1}{b_2})^2\sin{(p)}\cos{(p)}}{1+(\frac{b_1}{b_2})^2 \left(\cos{(p)}\right)^2}+b_3\\
    \Theta_\Gamma(p,\mathbf{b}) &= \frac{b_1\sin{(p)}}{1+(\frac{b_1}{b_2})^2 \left(\cos{(p)}\right)^2}+b_4.
\end{align}
The parameters $b_1$ and $b_2$ define the path's shape, while the parameters $b_3$ and $b_4$ define the path's mean azimuth and elevation.

To fit the kite's performance to the set of basis parameters, a recursive least squares (RLS) surface estimator is used. Using the estimated surface ($\tilde{J}(\mathbf{b}_{k})$), new basis parameters are computed at the end of each iteration using the following adaptation law:
\begin{equation}	\label{eqn:gradientBasedUpdateLaw}
\mathbf{b}_{k+1}=  \mathbf{b}_{k}+ K_{L} \nabla \tilde{J}(\mathbf{b}_{k}) + P_{n}.
\end{equation}
In this update law, $\mathbf{b}_{k}$ represents the basis parameters at iteration $k$, the matrix $K_{L}$ represents the learning gain and $\nabla \tilde{J}(\mathbf{b}_{\,k})$ is the gradient of the \emph{estimated} response surface, $\tilde{J}$, at the current basis parameters, $\mathbf{b}_{k}$. Additionally, the term $P_{n}$ is a perturbation term used to both explore the design space, driving the estimated response surface to the true response surface, in addition to pushing the update law out of local minima.  

In regard to the algorithm's convergence properties, readers are referred to \cite{cobb2019iterative} where under relatively restrictive assumptions on the mathematical nature of the performance index and variability in the environment, it was proven that the estimated response surface would converge to the true response surface, and the basis parameters would converge to a set containing the optimal basis parameters. Additionally, in \cite{cobb2022flexible}, it was demonstrated that even when the previously mentioned assumptions were loosened, convergence was still achieved.
\color{black}
\subsection{Lower-level Controller}
\textcolor{black}{The lower-level control system has two primary goals}:
\begin{enumerate}
    \item \textcolor{black}{Track the flight path prescribed by the higher level controller, which is ultimately achieved through control surface (aileron and rudder) actuation commands issued by the flight controller.}
    \item Strategically switch between spooling in and spooling out to ensure that the flight takes place at a relatively constant depth and flow speed range. This is done by controlling the spooling speed over the figure-8 lap and adjusting the kite's angle of attack (through use of the elevator) to ensure a high tension spool-out and a low tension spool-in. 
\end{enumerate}
This work leverages the control system designed in the co-authors' previous work \cite{reed2020optimal}, which uses a combination of a hierarchical flight controller and a winch controller in tandem, as shown in Fig. \ref{fig: control}. The control parameters are summarized in Table \ref{tab: ConParam}. While readers are referred to \cite{reed2020optimal} for the intricate details of the control system, it is summarized here for completeness.

\begin{table} [thb]
\centering
\caption{Control Parameters}
\renewcommand\arraystretch{1.2} 
\begin{tabular*}{1\columnwidth}{c @{\extracolsep{\fill}} c c } 
\hline
\hline
Parameter                & Description       & Unit\\ 
\hline

\textcolor{black}{$p$}                                & path position              & -              \\ 
$\gamma_{\mathrm{des}}$                     & desired velocity angle     & $rad$         \\ 
$\xi_{\mathrm{des}}$                        & desired tangent roll angle & $rad$         \\ 
$\mathbf{M_{des}}$                 & desired moment vector      & $Nm$          \\ 
$v_{\mathrm{spl}}$                          & spooling speed             & $m/s$         \\ 
$[\delta_e , \delta_a , \delta_r]$ & control surface deflection & $rad$         \\ 
\textit{\textbf{x}}                & position vector            & $m$           \\ 
\textit{\textbf{v}}                & velocity vector            & $m/s$         \\ 
$\mathbf{\omega}$                  & angular velocity vector    & $rad/s$       \\ 
$l_T$                              & un-spooled tether length   & $m$           \\ 
$[\phi , \theta , \psi ]$          & Euler angles               & $rad$         \\ 
\hline
\hline
\end{tabular*}
\label{tab: ConParam}
\end{table}

\subsubsection{Flight controller}
The hierarchical flight controller, shown within the dashed lines in the block diagram in Fig. \ref{fig: control}, has four levels. The prescribed  three-dimensional target cross-current path, \textcolor{black}{${\Gamma(p)}$}, is specified based on the Lemniscate of Booth, explained in \cite{rapp2019modular}. The path vector \textcolor{black}{${\Gamma(p)}$} is a function of the path position \textcolor{black}{$p$}, a scalar that varies between 0 and 2$\pi$ within a single figure-8 cycle, and describes the kite's location during the lap. \\
\indent The first level of the hierarchical controller uses the path vector along with the kite's position to determine a desired \emph{velocity angle}, $\gamma_{\mathrm{des}}$. This describes the desired direction of the kite's velocity vector. The second level uses the error between the desired and measured velocity angle to compute a desired tangent roll angle, $\xi_{\mathrm{des}}$, which is the angle between the kite body frame y axis, $\hat{y_k}$, and the so-called \emph{tangent plane}. The tangent plane is tangent to the surface of the sphere of radius $\left \| \emph{\textbf{x}} \right \|$ at the kite's instantaneous position. The third level of the hierarchical controller computes a desired moment vector, $\mathbf{M_{des}}$, based on the commanded and measured tangent roll angle. Finally, at the fourth level of the controller, the desired moments are then mapped to aileron and rudder deflections ($\delta_a$ and $\delta_r$, respectively) through a control allocation module.\\

\subsubsection{Winch controller}
The spooling motion is managed by the winch controller, which uses the kite's path position, $s$, to calculate the spooling speed $v_{\mathrm{spl}}$ and the elevator deflection $\delta_e$. This modulates the angle of attack to achieve high-tension spool-out motion and low-tension spool-in motion. The spooling speed is selected as one third of the flow velocity, following the optimization performed in \cite{loyd1980crosswind}. The elevator angle, $\delta_e$ is deflected to fixed values during the spool-out and spool-in phases that are known to drive the kite to high-tension and low-tension angles of attack, respectively. 

\subsection{Techno-economic metrics of interest} 
The effectiveness of an MHK system can be assessed through several techno-economic metrics. Some of the most prominently considered metrics include the following: 
\begin{enumerate}
    \item \emph{Generated power}: This refers to the net rated power output of the system. In this work, average and peak mechanical power are considered. 
    \item \emph{Structural mass}: This refers to the mass required by the support structures in the kite. For this work, the two most significant contributors, namely the structural masses of the wings and the fuselage, are considered. As structural mass is directly related to manufacturing cost, the design optimization aims to minimize it.  
    \item \emph{Peak-to-average power}: This refers to the ratio of the peak $P_{\mathrm{gen}}$ to the average $P_{\mathrm{gen}}$. 
    \item \emph{Levelized Cost Of Energy (LCOE)}: LCOE represents the average net present cost of generating power in the system through a specified period of time \cite{pawel2014cost}. 
\end{enumerate}

This works considers the first two techno-economic metrics. These two objectives, namely maximizing generated power and minimizing structural mass, are often considered concurrently in maximizing the power-to-mass ratio. In this work, we will consider the Pareto optimal formulation of minimizing mass subject to a power constraint, along with a dual-objective formulation that corresponds to the maximization of the \emph{exponentially weighted} power-to-mass ratio, given by:
\begin{equation} \label{eqn:weighted_power_to_mass}
R_{\mathrm{weighted}} = \frac{(P_{\mathrm{gen}})^{w}}{m_{\mathrm{wing}}}.
\end{equation}
\noindent This weighted formulation enables further generalization of the power-to-mass metric, allowing the designer to emphasize one aspect of performance over the other through the adjustment of $w$.

\section{OPTIMIZATION PROBLEM FORMULATIONS \label{sec: co-design}}
Ultimately, the techno-economic goals of the control-aware geometric and structural co-design formulation are to achieve high power and low mass, while satisfying geometric and structural constraints. In this work, we consider two optimization formulations for achieving these goals -- a Pareto optimal formulation and a dual-objective formulation.

The Pareto optimal formulation focuses on the minimization of structural wing mass, subject to an equality constraint on power generation as well as geometric and structural inequality constraints. The overall optimization problem is formulated as follows:
\begin{alignat}{2}
    &\underset{\textbf{u}}{\textup{minimize}}&&\quad m_{\mathrm{wing}} \label{eqn:genobj}\\
    &\text{subject to:} && \quad P_\mathrm{{gen}} = P_\mathrm{{req}} \label{eqn:Preq1}\\
    & && \quad C_w(\mathbf{u}) \leq 0 \label{eqn:wStructCon}\\
    & &&\quad C_f(\mathbf{u}) \leq 0 \label{eqn:fStructCon} \\
    & &&\quad m_{\mathrm{kite}}\leq \rho_w  V_{kite} \label{eqn:neutBuoyCon}\\
    & &&\quad \mathbf{u_{min,i} \leq u_i \leq u_{max,i}}, \vspace{2mm} \forall i.
\end{alignat}
\noindent where $\mathbf{u} = [\begin{array}{ccccc} s & AR &  N_\mathrm{{sp}} & t_\mathrm{{sp}} & t_{s,w} \end{array} ]^{T}$ is the vector of decision variables that are described in Table \ref{tab: decVar}, where $c$ and $t$ are the chord length and the thickness of the foil, respectively. Equation \eqref{eqn:Preq1} ensures that the required power generation is met. Inequalities \eqref{eqn:wStructCon} and \eqref{eqn:fStructCon} represent the structural constraints of the wing and fuselage, respectively, whereas \eqref{eqn:neutBuoyCon} ensures that the structural mass of the kite is less than the displaced mass of water, guaranteeing that neutral buoyancy can be achieved through sufficient ballast.

From an economic perspective, the Pareto optimal formulation is tailored toward situations where the target power output of a device is specified, and the remaining goal is to minimize the structural mass, which correlates highly with cost. By sweeping through a range of values of $P_{\mathrm{req}}$ in equation \eqref{eqn:Preq1}, this formulation can also be used to generate a Pareto front, as we in fact demonstrate in our results.


\begin{table} [thb]
\centering
\caption{\textcolor{black}{Optimization decision variables with bounds}}
\renewcommand\arraystretch{1.2} 
\begin{tabular*}{1\columnwidth}{c c c c } 
\hline
\hline
Variable                & Description & \textcolor{black}{Limits}       & Unit\\ 
\hline
\textcolor{black}{$s$}    & wing span               &\textcolor{black}{[7,10]}& $m$           \\ 
$AR$                    & wing aspect ratio       &\textcolor{black}{[4,12]}& -             \\ 
$N_{sp}$                & number of spars         &\textcolor{black}{[1,3]}& -             \\ 
$t_{sp}$                & thickness of spars      &\textcolor{black}{[0,20]}& $\%$ of $c$   \\ 
$t_{s,w}$               & thickness of wing shell &\textcolor{black}{[0,10]}& $\%$ of $t$   \\ 
$D$                     & diameter of fuselage    &\textcolor{black}{[0.4,0.8]}& $m$           \\ 
$L$                     & length of fuselage      &\textcolor{black}{[6,10]}& $m$           \\ 
$t_{s,f}$               & thickness of wing shell &\textcolor{black}{[0.5,10]}& $\%$ of $D$   \\ 
\hline
\hline
\end{tabular*}
\label{tab: decVar}
\end{table}

We also consider a dual-objective formulation, which allows for the explicit trade-off between mass and structural considerations and is formulated as follows: 
\begin{alignat}{2}
    &\underset{\textbf{u}}{\textup{maximize}}&&\quad       w \hspace{1mm}\mathrm{ln}(P_\mathrm{{gen}})-\mathrm{ln}(m_\mathrm{{wing}}) \label{eqn:genobj_dual}\\
    &\text{subject to:} && \quad P_\mathrm{{gen}} \geq P_\mathrm{{min}} \label{eqn:Preq1_dual}\\
    & && \quad C_w(\mathbf{u}) \leq 0 \label{eqn:wStructCon_dual}\\
    & &&\quad C_f(\mathbf{u}) \leq 0 \label{eqn:fStructCon_dual} \\
    & &&\quad m_{\mathrm{kite}}\leq \rho_w  V_{kite} \label{eqn:neutBuoyCon_dual}\\
    & &&\quad \mathbf{u_{min,i} \leq u_i \leq u_{max,i}}, \hspace{2mm} \forall i.
\end{alignat}

It can be easily shown, by taking the logarithm of both sides of equation \eqref{eqn:weighted_power_to_mass}, that the objective function in the above formulation is mathematically equivalent to the weighted power-to-mass ratio specified in equation \eqref{eqn:weighted_power_to_mass}. Consequently, this dual-objective formulation provides the designer with the explicit ability to specify the relative importance of power vs. structural mass in the optimization. Sweeping through the full range of $w$ also traces the convex portion of the Pareto front between the logarithmic structural mass and power plot. As a final observation, it is important to note that $P_{\mathrm{min}}$ in the dual-objective formulation (which replaces $P_{\mathrm{req}}$ from the Pareto optimal formulation) should be thought of as a \emph{minimum viable} power output, which will typically \emph{not} be satisfied at equality, since there can exist considerable economies of scale from increases in the power output. For example, consider the case where $w = 1$, $P_\mathrm{min} = 100 kW$, and two designs are possible: (i) a design in which $m_{\mathrm{wing}} = 700 kg$ and $P_{\mathrm{gen}} = 100 kW$ and (ii) a design in which $m_{\mathrm{wing}} = 1000 kg$ and $P_{\mathrm{gen}} = 200 kW$. The second design, for which the minimum viable power constraint is not met at equality, is superior according to the dual-objective formulation.

\section{Individual Optimization Modules}
The overall co-design formulation detailed in Section \ref{sec: soltn} makes use of four constituent modules, which are described below and detailed in this section:
\begin{itemize}
    \item[--] A \emph{Steady flight optimization tool (SFOT)}, which selects the geometric properties of the wings and the stabilizers to minimize a surrogate for wing mass, subject to geometric and performance constraints; 
    \item[--] A \emph{Structural wing design tool (SWDT)}, which selects span and shell properties of the wing frame to minimize wing mass, subject to wing tip deflection constraints; 
    \item[--]A \emph{Structural fuselage design tool (SFDT)}, which selects shell thickness to minimize fuselage mass, subject to hoop stress, shear stress and buckling constraints. 
    \item[--] A \emph{closed-loop flight efficiency map}, which approximates the ratio of achieved peak mechanical power output to the theoretical upper limits predicted by the quasi-steady analysis in \cite{loyd1980crosswind}.
\end{itemize}
Note that each of the first three modules can be used as a stand-alone program, in addition to being usable as part of the overall co-design formulation. The flight efficiency map ultimately eliminates the need for controller re-tuning and dynamic simulations for every iteration of plant design within the overall co-design process.

\subsubsection{Steady flight optimization tool \label{sec: sfot}}
The steady flight optimization tool seeks to obtain the most compact wing design that can produce the required power for a rated flow speed. This is done is by minimizing the wing volume while meeting the performance and geometric constraints: 

\begin{alignat}{2}
    &\underset{\textbf{u}}{\textup{minimize}}&&\quad f_\mathrm{{surr}}(\mathbf{u_{SF}}) \label{eqn:SFOTobj}\\
    &\text{subject to:} && \quad \max_{s,AR}  \left (\frac{2}{27} \right)\frac{\eta \rho_w v^3 s^2}{AR^2}\left ( \max_{\alpha} \frac{C_L^3}{C_D^2} \right ) \geq P_{\mathrm{req}} \label{eqn:perfReqIneq}\\
    & &&\quad \mathbf{u_{SF,min,i} \leq u_{SF,i} \leq u_{SF,max,i}} \hspace{2mm} \forall i.
\end{alignat}

\noindent where $\mathbf{u_{SF}} = [\begin{array}{cc} s & AR \end{array}]^{T}$ are the decision variables, $f_{surr}(\mathbf{u_{SF}})$ is the objective function that acts as a surrogate to minimizing wing mass (further detailed in Section 4), $\eta$ is the flight efficiency (explained in Section \ref{sec: eff}), $v$ is the rated flow speed, $C_L$ and $C_D$ are the lift and drag coefficients of the whole kite, and $\alpha$ is the angle of attack. The generated mechanical power estimate shown in \eqref{eqn:perfReqIneq} is based on the seminal work of Miles Loyd, as detailed in \cite{loyd1980crosswind} and referenced throughout the AWE and MHK kite literature. 

\subsubsection{Closed-loop flight efficiency map \label{sec: eff}}
The value of $P_{\mathrm{gen}}$ estimated in \eqref{eqn:perfReqIneq} is based on steady cross-current flight estimates. This is an idealized estimate that does not account for transient effects associated with the kite's acceleration and deceleration within a figure-8 cycle. Furthermore, the idealized estimate in \eqref{eqn:perfReqIneq} assumes that the kite can be spooled in infinitely fast, under zero tension.\\
\indent The closeness of the kite's actual power output to steady cross-current flight predictions depends significantly on the kite's closed-loop control performance, where the achievable performance depends on geometric kite parameters (especially the wing parameters). For the underwater kite system, the controller has a largely unidirectional coupling with the plant, wherein the plant decision variables $\mathbf{u_p}$ significantly impact the optimal achievable control performance, but knowledge of the controller is significantly less important in optimizing the plant. This dependence was studied through a sensitivity analysis, whereby the influence of key plant parameters within $\mathbf{u_p}$ on flight efficiency was characterized. To account for the impact of design parameters on achievable closed-loop flight performance, a flight efficiency term can be defined as:
\textcolor{black}{
\begin{equation}
\eta = \frac{P^*_{\mathrm{gen}}}{P_\mathrm{{ideal}}}
\end{equation}}
\textcolor{black}{\noindent where $P^*_{\mathrm{gen}}$ is the optimal simulated peak generated mechanical power (which is equal to the converged power resulting from the economic ILC-based flight control optimization), and $P_\mathrm{{ideal}}$ is the steady flight power calculated in \eqref{eqn:perfReqIneq} with $\eta$ = 1.}

\textcolor{black}{Several kites with varying wing designs were simulated to perform the sensitivity analysis. For each candidate kite design, mass and inertia properties were computed based on the kite's geometry based on the displaced volume of fluid. Furthermore, the control surfaces and fuselage were scaled based on the kite's size.}

\begin{figure}[h] 
\begin{center}
  	\includegraphics[width = \columnwidth]{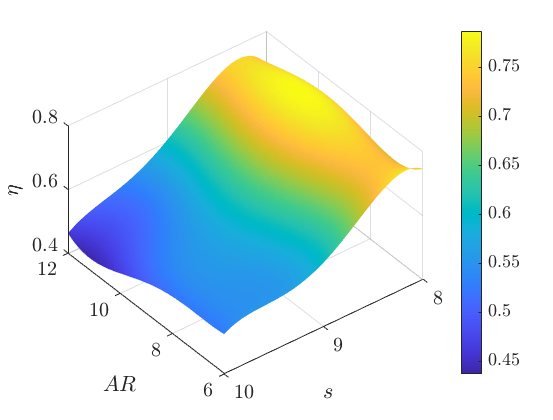}
	\caption{\textcolor{black}{Results of sensitivity analysis: A flight efficiency map that relates the optimized dynamically simulated power generated $P^*_{\mathrm{gen}}$ to the theoretically calculated and optimized power $P_\mathrm{{ideal}}$ calculated in \eqref{eqn:perfReqIneq}}. \label{fig: effmap}}  
\end{center}
\end{figure}

The flight performance of the kite depends largely on the wing definition, and consequently, the efficiency map was characterized as a function of \textcolor{black}{$s$} and $AR$. For each simulation, the efficiency was recorded and mapped to a point in the \textcolor{black}{$s-AR$} design space. A surface was then fit to obtain the closed-loop flight efficiency as a function of the plant decision variables, $\eta(s,AR)$. This function acts as a \emph{control proxy function}, a concept initially introduced to the co-design literature in \cite{peters2011control}, which allows the plant design optimization to be formulated without explicit consideration of the controller (and without the need to re-optimize and re-simulate the controller for each plant design variant). The efficiency map obtained from the sensitivity analysis is illustrated in Fig. \ref{fig: effmap}. 

\subsubsection{Structural wing design tool}
The lifting loads acting on the wing produce a large bending moment about the chord-wise neutral axis. The wing structural design tool minimizes the structural support mass required to support the wing. The structure is designed as a combination of spars and a shell. \\
\indent The SWDT solves the mixed-integer constrained optimization problem, formulated as: 
\begin{alignat}{2}
    &\underset{\textbf{u}}{\textup{minimize}}&&\quad m_\mathrm{{wing}}(\mathbf{u_{WD}}) = \rho_m s A_\mathrm{{wing}}(\mathbf{u_{WD}}) \label{eqn:SWDTobj}\\
    &\text{subject to:} && \quad I_\mathrm{{wing}} \geq I_\mathrm{{req}}\mid \delta_\mathrm{{max}} \label{eqn:Ireq}\\
    & &&\quad \mathbf{u_{WD,min,i} \leq u_{WD,i} \leq u_{WD,max,i}} \hspace{2mm} \forall i,
\end{alignat}

\noindent where $\rho_m$ is the material density, $A_\mathrm{{wing}}$ is the total cross-sectional area, and $\mathbf{u_{WD}} = [\begin{array}{ccc} N_\mathrm{{sp}} & t_\mathrm{{sp}} & t_{s,w} \end{array}]^{T}$ are the decision variables. The bounds on the decision variables are as follows: 
\begin{itemize}
\item[--] $N_\mathrm{{sp}}$ can take integer values among $N_\mathrm{{sp}} \in (1,2,3)$ ;
\item[--] Bounds on both $t_\mathrm{{sp}}$ and $t_{\mathrm{s,w}}$ are linear functions of the chord length $c$ of the wing, given by: $t_{\mathrm{sp,min}}(c) \leq t_\mathrm{{sp}} \leq t_{\mathrm{sp,min}}(c)$, $t_{\mathrm{s,w,min}}(c) \leq t_{\mathrm{s,w}} \leq t_{\mathrm{s,w,min}}(c)$.
\end{itemize}

\subsubsection{Structural fuselage design tool \label{sec: SFDT}}
The structural fuselage design tool minimizes the mass of the fuselage while meeting several structural constraints. A simplifying assumption is made to design the fuselage as a cylindrical shell. The SFDT solves the optimization problem formulated as: 

\begin{alignat}{2}
    &\underset{\textbf{u}}{\textup{minimize}}&&\quad m_\mathrm{{fuse}}(\mathbf{u_{FD}}) = \rho_m A_{\mathrm{fuse}}(\mathbf{u_{FD}}) L \label{eqn:SFDTobj}\\
    &\text{subject to:} \!\!&&\quad \frac{\sum F_{z}}{t_{s,f}L}\leq \zeta \sigma_{\mathrm{yield}} \label{eqn:shear}\\
    & &&\quad \frac{PD}{2t_{s,f}}\leq\zeta \sigma_{0.5}\label{eqn:hoop}\\
    & &&\quad \frac{\left | M_{\mathrm{max}} \right |}{S(\mathbf{u_{FD}})}\leq \zeta \sigma_{\mathrm{yield}}\label{eqn:buck}\\
    & &&\quad \mathbf{u_{FD,min,i} \leq u_{FD,i} \leq u_{FD,max,i}}, \hspace{2mm} \forall i,
\end{alignat}

\noindent where $A_{\mathrm{fuse}}$ is the cross-sectional area of the fuselage, and $\mathbf{u_{FD}} = [\begin{array}{ccc} D & L & t_{s,f} \end{array}]^{T}$ are the decision variables. Equations \eqref{eqn:shear}, \eqref{eqn:hoop}, and \eqref{eqn:buck}, respectively place constraints on the shear stress (induced due to tangential loads), hoop stress (induced due to pressure difference $P$; designed for 0.5 $\%$ elongation), and buckling loads due bending moments ($M_{\mathrm{max}}$ is the maximum buckling moment, and ${S(\mathbf{u_{FD}})}$ is the section modulus) \cite{roylance2001pressure}. 

\begin{itemize}
    \item[--] \textit{Shear stress:} Equation \eqref{eqn:shear} models the shear behaviour due tangential loads at the wing and stabilizer attachment points. $F_z$ are the loads in the direction normal to the wing, $\zeta$ is the factor of safety, and the $\sigma_{\mathrm{yield}}$ is the yield stress of the material. 
    \item[--] \textit{Hoop stress:} The fuselage of the kite is assumed to be a thin-walled pressure vessel. The difference in the external and internal pressures, $P$, causes circumferential stresses. The variable $\sigma_{0.5}$ is the stress at 0.5 $\%$ elongation. 
    \item[--] \textit{Buckling:} The lift forces of the wings and the horizontal stabilizer induce buckling loads about the tether attachment point. $\left | M_{\mathrm{max}} \right |$ is the maximum buckling moment calculated for a set of hydrodynamic forces, and ${S(\mathbf{u})}$ is the section modulus of the fuselage. 
\end{itemize}


\section{Integrated Co-Design Solution Approach \label{sec: soltn}}
In this section, we present two candidate approaches for fusing the previously described individual optimization/analysis modules into an integrated Pareto optimal co-design solution. We also present a simultaneous optimization approach for addressing the dual-objective problem. Graphical depictions of these solution techniques are provided in Fig. \ref{fig: compStrat}, and detailed descriptions of each approach are provided in the following subsections.

\subsection{Pareto optimal co-design via the nested-sequential approach} This approach, as originally discussed in our conference publication \cite{naikfused}, involves a sequential execution of the SFOT and SWDT, along with a calculation of required fuselage thickness, nested inside of an outer loop that iterates on the fuselage diameter ($D$) and length ($L$). Because the span ($s$) and aspect ratio ($AR$) alone (which constitute the decision variables for the SFOT) are not sufficient to compute the kite's mass (which is the variable to be minimized in the Pareto optimal formulation), the SFOT must rely on a \emph{surrogate} objective function, $f_{\mathrm{surr}}$, which (i) is closely correlated with kite mass and (ii) is strictly a function of $s$ and $AR$. In this work, two candidate surrogate objectives are considered for the SFOT: (i) wing volume ($V_{\mathrm{wing}}$ and (ii) span ($s$). In the event that span, which is also a decision variable, is used as the surrogate objective, the SFOT boils down to the finding the minimum value of $s$ for which the power equality constraint, $P = P_{\mathrm{req}}$, remains feasible for an admissible value of $AR$.

Coupling between the geometric design and required mass to achieve structural constraints within the SWDT leads, in general, to sub-optimality of the nested-sequential approach. Specifically, there exists a unidirectional coupling between SFOT and SWDT -- SFOT chooses $s$ and $AR$, which directly affects the structural optimization, as it determines the chord length of the wing. Additionally, the hydrodynamic forces on the wing are determined by SFOT, which directly affect the wing deflection constraint in SWDT. On the other hand, the wing structural design does not affect the steady flight performance, except in the case where no structural design within the bounds of the structural variables satisfies the structural constraints. This is observed when the wing design has a large wing span, $s$, and large aspect ratio, $AR$ (and consequently a small chord length $c$). The results of Section 6 will further investigate the impact of coupling and choice of $f_{\mathrm{surr}}$ on the performance of the nested-sequential approach.  

\subsection{Pareto optimal co-design via the fully nested approach} 
\begin{figure}[h] 
\begin{center}
  	\includegraphics[width = \columnwidth]{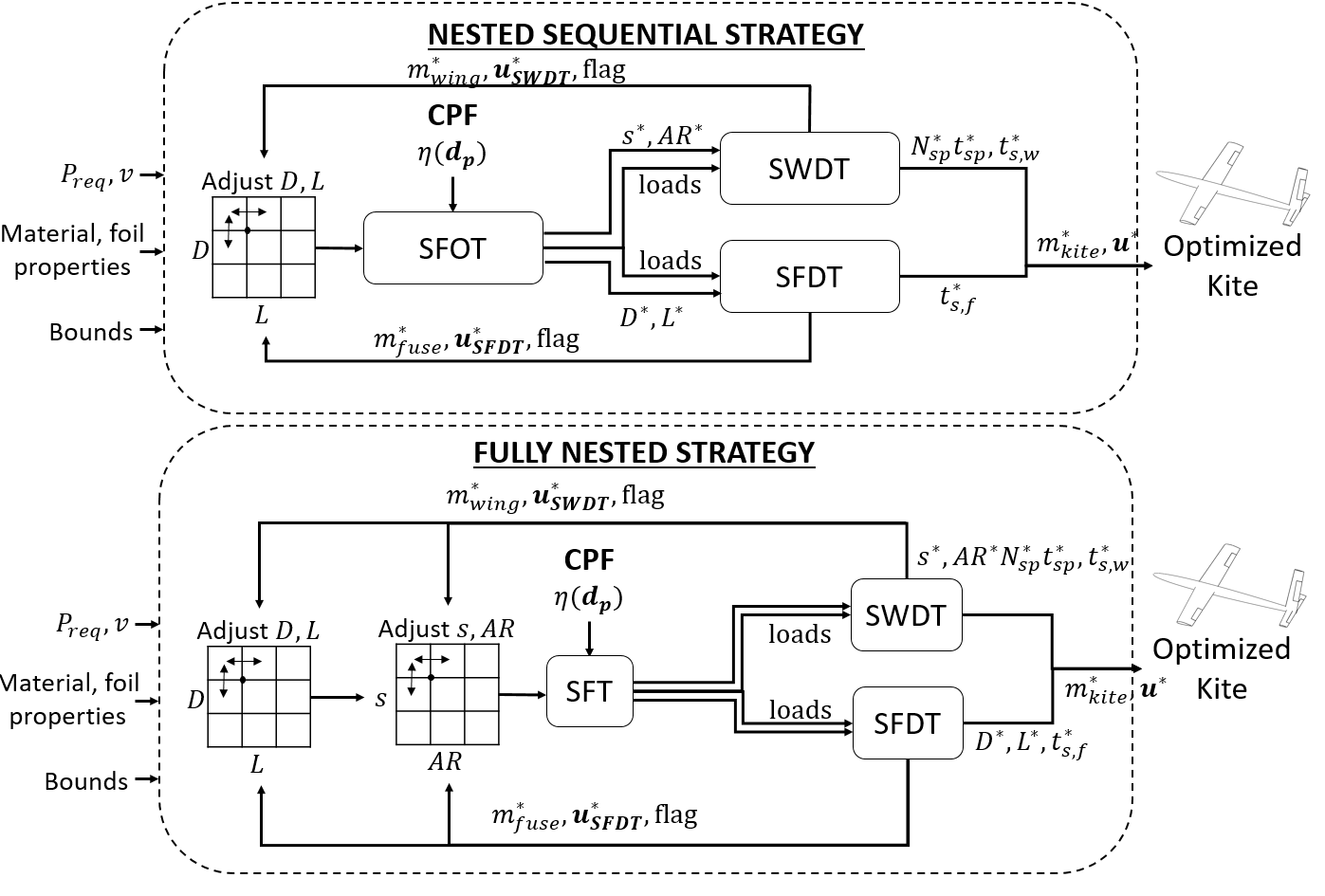}
	\caption{Block diagrams for the two solution strategies introduced in the paper: (i) Sequential nested strategy and (ii) Fully nested strategy. \label{fig: compStrat}}  
\end{center}
\end{figure}

In this approach, the SWDT and a calculation of the required fuselage thickness are nested inside of an outer loop that iterates on the wing geometry ($s$ and $AR$), as well as the fuselage diameter ($D$) and length ($L$). Within the outer loop, only the combinations of geometric variables satisfying the power equality constraint, $P = P_{\mathrm{req}}$, are chosen (performed by the steady flight tool (SFT) as shown in Fig.\ref{fig: compStrat}, which does not perform an optimization but chooses sets of decision variables that satisfy constraints). It is important to note that the wing structural parameters and fuselage thickness do not impact the satisfaction of the power equality constraint, allowing for the consideration of the constraint within the outer loop. While less computationally efficient than the nested-sequential strategy, the fully nested approach enables convergence to a globally optimal design.

\subsection{Dual-objective co-design via a simultaneous solution \label{sec:dualObj}}
Both the nested-sequential and fully nested strategies work through minimization of a mass metric, subject to an equality constraint on power. In the sequential strategy, this is needed for the partitioning of the SFOT and SWDT (in addition to requiring a surrogate objective for the SFOT), whereas the nested strategy requires the equality constraint on power in order to restrict the consideration of the outer-loop design space to geometric variables that satisfy the power constraint. Because neither co-design formulation is tailored to a dual-objective formulation, we rely on a simultaneous strategy when considering the dual-objective formulation.

As indicated in Fig.\ref{fig: compStrat}, the simultaneous optimization represents a mixed integer formulation, consisting of a substantial number of continuous decision variables (wing span and aspect ratio; fuselage shell thickness, diameter and length; wing spar thickness and shell thickness), along with an integer variable (number of spars). In this work, the simultaneous optimization has been solved through the use of a genetic algorithm (GA). Due to the relatively complex (non-convex with multiple local minima) nature of the structural optimization tools, GA was chosen as the solution technique to obtain a global solution. The GA was implemented with a population size of 200 and an elite count of 20. The efficacy of the formulation is shown in Section \ref{sec: ParetoComp}, where the results of the simultaneous solution using GA are compared to the Pareto-optimal solution obtained using the fully nested formulation. 


\section{Results}

In this Section, we examine the kite designs and closed-loop performance that results from the solution formulations described in Section \ref{sec: soltn}. For this effort, we focus on a case study in a constant flow environment with a flow speed $v = 1.5 m/s$. The dynamic model and controller presented in \cite{reed2020optimal} was used to simulate cross-current flight. The simulation parameters used are summarized in Table \ref{tab: SimParam}. 

\begin{table} [thb]
\centering
\caption{Simulation parameters}
\renewcommand\arraystretch{1.2} 
\begin{tabular*}{1\columnwidth}{c @{\extracolsep{\fill}} c c c } 
\hline
\hline
\textbf{Property} & \textbf{Description}                        & \textbf{Value} & \textbf{Unit} \\ 
\hline
$\delta_\mathrm{{max}}$    & max. wingtip deflection & 5$\%$              & -               \\ 
$\sigma_{\mathrm{yield}}$  & material yield   strength               & 2.70E+08       & $Pa$          \\ 
$E$               & material Young's   modulus               & 6.89E+10       & $Pa$          \\ 
$P$               & pressure difference                         & 2.50E+03       & $Pa$          \\ 
$\gamma$          & lift   curve multiplier                     & 9.60E-01       & -             \\ 
$e_L$             & Oswald   lift efficiency                    & 7.60E-01       & -             \\ 
$e_D$             & Oswald   drag efficiency                    & 9.20E-01       & -             \\ 
$C_{L,0}$         & zero   AoA lift coefficient                 & 1.60E-01       & -             \\ 
$C_{L,x}$         & lift   at minimum drag                      & 2.00E-02       & -             \\ 
$K_{\mathrm{visc}}$        & viscous drag factor                         & 3.00E-02       & -             \\ 
$C_{D,0}$         & drag   at zero lift                         & 6.50E-03       & -             \\ 
$v$          & flow speed of water                            & 1.50E+01       & $m/s$      \\ 
$\rho_w$          & water density                           & 1.00E+03       & $kg/m^3$      \\ 
$\rho_{m}$       & material density                         & 2.70E+03       & $kg/m^3$      \\ 
\textcolor{black}{$l_{T}$}       &  \textcolor{black}{un-spooled tether length}                         & \textcolor{black}{1.250E+02}       & \textcolor{black}{$m$}      \\ 
\textcolor{black}{$r_{thr}$}       & \textcolor{black}{radius of tether}                         & \textcolor{black}{5.00E-02}       & \textcolor{black}{$m$}      \\ 
\textcolor{black}{$C_{D,thr}$}       & \textcolor{black}{tether drag coefficient}                         & \textcolor{black}{1.000E+01}       & \textcolor{black}{-} \\ 
\hline
\hline
\end{tabular*}
\label{tab: SimParam}
\end{table}

\subsection{Comparison of Pareto optimal and dual-objective optimizations \label{sec: ParetoComp}}
We first compare the resulting kite mass and power output resulting from a fully nested Pareto optimal formulation against those resulting from the dual-objective formulation using simultaneous optimization. Because both the fully nested and simultaneous formulations enable convergence to global optima of their respective objective functions, they enable a fair and meaningful comparison between the Pareto optimal and dual-objective optimization problems.

\begin{figure}[h] 
\begin{center}
  	\includegraphics[width = \columnwidth]{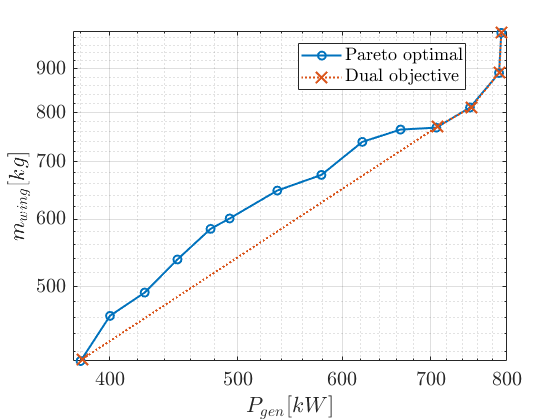}
	\caption{\textcolor{black}{Results comparing the dual-objective formulation with the Pareto front resulting from the Pareto optimal formulation. It can be seen that the dual-objective result traces the convex portions of the Pareto front.} \label{fig: pareto}}  
\end{center}
\end{figure}

Fig.\ref{fig: pareto} shows two data sets. The first, shown in the blue curve (and corresponding circles), is the result of a sweep of Pareto optimal designs, where the simultaneous optimization formulation was used to minimize $m_{\mathrm{wing}}$, subject to a range of power constraints, sweeping from \textcolor{black}{350 $kW$ to 775 $kW$}. The second data set characterizes the results of the dual-objective framework, using the fully nested co-design approach. Here, the weighting variable $w$ was swept from 0.01 (prioritizing power more than mass) to 100 (prioritizing mass more than power), with the resulting performance indicated in Fig. \ref{fig: pareto} by red x's.

Several important insights can be gleaned from Fig.\ref{fig: pareto}. First, a vertical asymptote appears to exist around \textcolor{black}{$P_{\mathrm{gen}} = 775 kW$}, at which point no further increase in power output is possible. Indeed, this arises due to physical limits on the kite's span. Secondly, the results of the dual-objective optimization follow the same pattern as those of the Pareto optimization for large power outputs (beyond approximately \textcolor{black}{700 $kW$}) and at the low end of the power spectrum (around \textcolor{black}{375 $kW$}) but do not appear to allow for designs that generate intermediate amounts of power. This is indeed explainable due to the concave nature of the Pareto front between \textcolor{black}{375 $kW$ and 700 $kW$}. In fact, the results of the dual-objective formulation trace out the convex portion of the Pareto front in a log-log space.

\begin{figure*}[h] 
\begin{center}
  	\includegraphics[width = 0.875\linewidth]{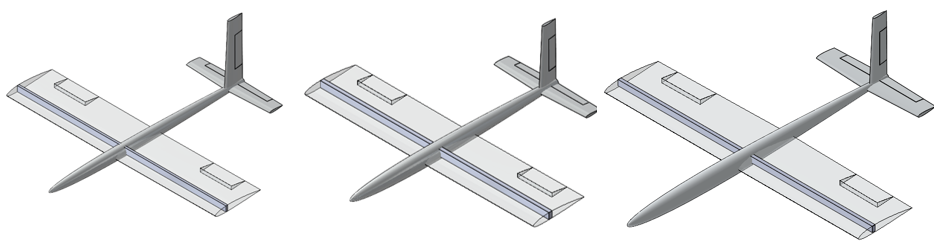}
	\caption{\textcolor{black}{CAD comparison between the mass-driven (left), intermediate (middle), and power-driven (right) designs.} \label{fig: cadComp}}  
\end{center}
\end{figure*}

In addition to the performance comparisons from Fig.\ref{fig: pareto}, it is also instructive to examine the physical geometric and structural designs corresponding to different points along the Pareto front. To that end, the design specifications of three kite designs corresponding to points along the Pareto front are presented in Table \ref{tab: DesParam}. Furthermore, the physical geometries and wing structural cross sections are shown in Fig. \ref{fig: cadComp}. As the minimum power constraint is increased (corresponding to a power-driven optimization), the wing span ($s$) increases to its upper limits, and the aspect ratio ($AR$) decreases to maximize planform area. This increased wing span is associated with diminishing returns in terms of an unweighted mass-to-power ratio, but the increased span is essential when either the minimum required power is set to a very high value (in the Pareto optimal formulation) or the weight on mass is taken to be very large in the dual-objective formulation ($w >> 1$). Furthermore, the smaller aspect ratio ($AR$) is associated with decreased power per unit mass, due to the decreased wing efficiency; however, absolute power can be increased, to a point, through a decrease in $AR$. Unsurprisingly, the spar structure must also be thickened for higher-power wings, in order to accommodate the increased loading.

\begin{table} [thb]
\centering
\caption{\textcolor{black}{Comparison of kite designs: Designs A,B and C represent mass-driven, intermediate and power-driven designs respectively.}}
\renewcommand\arraystretch{1.2} 
\begin{tabular*}{1\columnwidth}{|>{\color{black}}c|>{\color{black}}c|>{\color{black}}c|>{\color{black}}c|>{\color{black}}c|} 
\hline
\hline
Property & Unit                        & Design A & Design B & Design C\\ 
\hline
$s$                                                             & $m$                                                        & 7.08                                 & 8.51                                      & 9.98                                                                       \\ 
$AR$                                                            & -                                                          & 6.50                                 & 6.00                                        & 4.70                                                                       \\ 
$t_{sp}$                                                        & $\%$ of $c$                                                & 11.2                                 & 13.9                                              & 12.8                                                                       \\ 
$t_{s,w}$                                                       & $\%$ of $t$                                                & 0.80                                 & 0.78                                     & 0.64                                                                    \\ 
$D$                                                             & $m$                                                        & 0.51                                  & 0.59                                      & 0.70                                                                       \\ 
$L$                                                             & $m$                                                        & 6.4                                  & 7.0                                      & 7.5                                                                    \\ 
$t_{s,f}$                                                       & $\%$ of $D$                                                & 1.3                                  & 1.8                                     & 1.5                                                                     \\ 
$m_{fuse}$                                                      & $kg$                                                       & 231.1                                & 387.8
& 532.5\\ 
$m_{wing}$                                                      & $kg$                                                       & 404.8                               & 628.7                                       & 891.2                                                                     \\ 
$m_{kite}$                                                      & $kg$                                                       & 635.9                                & 1016.5& 1423.7                                                                     \\ 
$P_{gen}$                                                       & $kW$                                                       & 371.2                                & 548.3                                      & 802.2                                                                      \\ 
$P_{avg}$                                                       & $kW$                                                       & 95.7                                & 134.3                                      & 191.2                                                                      \\ 
\hline
\hline
\end{tabular*}
\label{tab: DesParam}
\end{table}

The CAD of the aforementioned designs are shown in Fig. \ref{fig: cadComp}. This illustrates the comparison between the sizes of the wings, the fuselages and the resulting structural frames.

\begin{figure}[tbh] 
\begin{center}
  	\includegraphics[width = \columnwidth]{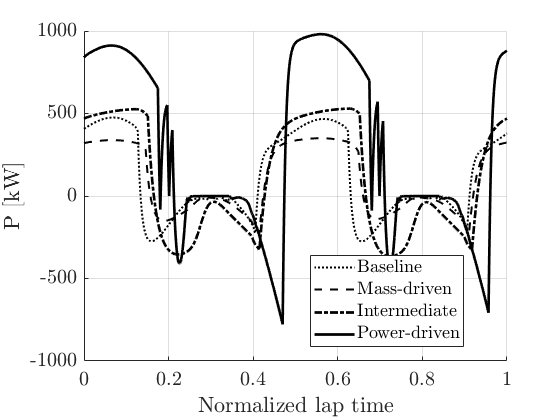}
	\caption{\textcolor{black}{Flight performance comparison: The power outputs obtained through simulation of the three kite designs with the closed-loop flight controller, plotted against normalized lap time for a duration of one lap. The flight performance of the baseline is also compared with the kites, which lies in between the mass driven (least $P_{gen}$) and the intermediate design. \label{fig: PcompCL}}}  
\end{center}
\end{figure}

Finally, given that the kite exhibits dynamic motion, it is instructive to examine the closed-loop flight performance obtained with each of the aforementioned three designs (mass-driven, intermediate, and power-driven, respectively labeled as Designs A, B and C). The Fig.\ref{fig: PcompCL} compares the simulated closed-loop flight performance in each case, over a representative figure-8 cycle, which consists of two spool-in phases and two spool-out phases. It is important to note that the bulkier power-driven kite has significantly larger mass moments of inertia (driven by significantly larger structural mass), requiring a larger figure-8 path to be able to traverse the turns. The larger dip in the power profile for the power-driven kite shows that it also uses more power in the spool-in phase. Figure \ref{fig: PcompCL} also compares the performance of the baseline kite presented in one of the co-author's previous papers \cite{reed2020optimal}. As noted in our previous work \cite{naikfused}, no feasible structure within the specified bounds can be designed for the baseline due to significantly larger bending moments acting on the wing with high $AR$ and large span $s$. However, for comparison, the baseline and the intermediate designs were simulated in a flow environment of 1.5 m/s. The intermediate design generated a \textcolor{black}{$P_{avg} = 58.23 kW$} and \textcolor{black}{$P_{peak} = 234.2 kW$}. This resulted in power-to-mass ratios of \textcolor{black}{$P_{avg}/m_{kite} = 0.057 kW/kg$} and \textcolor{black}{$P_{peak}/m_{kite} = 0.23 kW/kg$} as compared
the baseline power-to-mass ratios, $P_{avg}/m_{kite} = 0.012 kW/kg$ and $P_{peak}/m_{kite} = 0.062 kW/kg$. Thus, a three-fold improvement in the power-to-mass ratio from the baseline design was observed. 

\subsection{Comparison of fully nested and nested-sequential strategies for Pareto optimal design}
As noted earlier, the nested-sequential optimization approach, which requires the SFOT to rely on a surrogate measure for wing mass (since wing mass depends on structural decision variables outside of the SFOT), will in general result in sub-optimality due to the coupling between the geometric and structural variables. The level of sub-optimality will, in general, depend on the surrogate metric ($f_{\mathrm{surr}}$) used by the SFOT in place of structural wing mass.

\begin{figure}[tbh] 
\begin{center}
  	\includegraphics[width = \columnwidth]{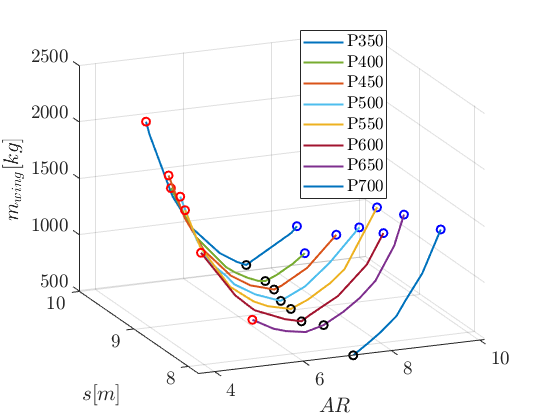}
	\caption{\textcolor{black}{Results comparing the minima using the surrogate objectives in the sequential approach and the nested approach. For each curve, the red marker denotes the solution using the sequential-nested strategy with $f_{\mathrm{surr}} = s$, the blue marker denotes the solution using the sequential-nested strategy with $f_{\mathrm{surr}} = V_{wing}$, and the black marker denotes the solution using the fully nested strategy.} \label{fig: surveMin}}     
\end{center}
\end{figure}

To understand the extent of that sub-optimality, a fully nested optimization was performed over a range of power (equality) constraints, varying from \textcolor{black}{$P_{req} = 350 kW$ to $700 kW$}. The curves generated by running the wing design optimization for all wing sets of wing spans and aspect ratio that meet the performance constraint, are shown in Fig.\ref{fig: surveMin}. To compare the strategies, the minima obtained through a fully nested strategy are compared against the minima obtained under the nested-sequential approach, considering two candidate surrogate metrics ($f_{\mathrm{surr}}$): (i) wing span ($s$) and (ii) wing volume (proportional to $\frac{s^{3}}{AR^{2}}$). Fig.\ref{fig: surveMin} shows the results of this study.

It can be seen that the nested-sequential strategy leads to some level of sub-optimality over the whole range of power requirements, where the level of sub-optimality is particularly pronounced in many instances where $f_{\mathrm{surr}} = s$ is used. To better understand the reason for these results, it is important to understand the physics behind the two different surrogate metrics: 
\subsubsection{Nested-sequential strategy using $f_{\mathrm{surr}} = s$}
In this case, the kite design with the smallest span that meets the power requirements is chosen. In such cases, a large chord length, or small wing $AR$, is required to provide the necessary wing planform area for satisfying the performance constraint (eq. \eqref{eqn:perfReqIneq}). This might not always be desirable, as the wing may satisfy the bending stiffness required (from eq. \eqref{eqn:fStructCon}) and converge to the lower bounds of the structural decision variables, but as they vary linearly with the chord length $c$, which is large in such designs, this strategy results in an over-design. 
\subsubsection{Sequential strategy using $f_{\mathrm{surr}} =  V_{\mathrm{wing}}$}
For a given foil geometry (which is assumed to be that of \textcolor{black}{NACA 2412} in this work), which fixes the wing's thickness-to-chord ratio, the wing volume is mathematically proportional to $V_{wing} = s^3/AR^2$. Thus, minimizing $V_{\mathrm{wing}}$ directly incentivizes maximizing the aspect ratio ($AR$). Thus, the wings resulting from such an optimization have very high aspect ratios, which are obtained through a large span ($s$). A large span is efficient from the standpoint of lift-induced drag, which helps to reduce the amount of total wing area required to achieve a specified level of power. However, large spans also result in higher bending stiffness requirements for the structural frame of the wing, which results in a increased wing structural mass. 

\section{Conclusions}
This work presented Pareto optimal and dual-objective formulations for the control-aware optimization of the combined geometric and structural properties of an underwater kite. The resulting designs are capable of minimizing required structural mass subject to an equality constraint on power and maximizing a weighted power-to-mass ratio. Several co-design formulations were presented for performing these optimizations, including a nested-sequential, fully nested, and simultaneous approach. In the latter two approaches, the full optimization makes use of individual geometric and structural optimization modules, which also can perform stand-alone optimizations. The application of these optimization techniques resulted in a three-fold improvement relative to an earlier baseline kite design used by the research group, along with significant insights regarding the trade-offs between structural, hydrodynamic, and closed-loop flight performance considerations. 

\textcolor{black}{Future work will focus on consideration of transient performance, in addition to the cycle-averaged power and peak structural loading considered in the present work. Most notably, this will involve augmentation of the structural tool to consider cyclic fatigue loading over the course of each lap, which will first be considered in steady flow conditions, followed by consideration in turbulent scenarios. Additionally, owing to the fact that true environmental characteristics are time-varying and site-dependent, future work will incorporate real time-varying flow characteristics into the optimization framework, in addition to ultimately considering site selection within the co-design formulation.}



\section*{Funding Data}
This work was supported by the US Department of Energy, under the award entitled ``Device Design and Robust Period Motion Control of an Ocean Kite System for Hydrokinetic Energy Harvesting" (US Department of Energy Award No. DE-EE0008635), and the National Science Foundation, under the award entitled ``CAREER: Efficient Experimental Optimization for High-Performance Airborne Wind Energy Systems" (Award No. 1914495). The work was also supported by the North Carolina Renewable Ocean Energy Program.

%
% ...

\bibliographystyle{asmems4}

\bibliography{asme2ej}

\end{document}